Rainer Burghardt

# A Kerr Interior

An exact solution of the Einstein field equations is proposed which represents a differentially rotating fluid. As this solution matches the exterior Kerr solution and reduces to the Schwarzschild interior solution by setting the rotational parameter to zero, it could serve as Kerr interior.

## 1. Introduction

In the last decades many searchers have dealt with the construction of a solution of the Einstein field equations for a rotating source that matches the Kerr solution. Approximate solutions and trial solutions have been found [1-22]. We want to propose another solution by means of geometrical methods.

In Sec. 2 of this paper we present an interior for the Kerr metric based on a differentially rotating fluid source. This exact solution of the Einstein field equations matches the Kerr vacuum solution at a boundary surface of an elliptical shape. Setting the rotational parameter to zero it has as static limit the Schwarzschild interior solution. In Sec. 3 we study a static seed metric for investigating the geometrical background of the model. We set up the field equations and calculate the stress-energy tensor. In Sec. 4 we implement the rotation by an intrinsic transformation operating on the 4-bein fields and we calculate the rotational and centrifugal forces. We set up the field equations for the rotating system and we calculate the stress-energy tensor of the rotating masses.

## 2.   The rotating metric

Firstly, we write down the line element of the interior region and we define all the quantities we will use throughout the paper. Using the elliptical Boyer-Lindquist co-ordinates the line element reads as

$$ds^2 = dx^{1^2} + dx^{2^2} + \left[ \alpha_R dx^3 + i\alpha_R \omega \sigma dx^4 \right]^2 + a_T^2 \left[ -i\alpha_R \omega \sigma dx^3 + \alpha_R dx^4 \right]^2, \quad (2.1)$$

wherein

$$dx^1 = \alpha_I a_R dr, \quad dx^2 = \Lambda d\vartheta, \quad dx^3 = \sigma d\varphi, \quad dx^4 = idt$$

$$\alpha_I = \frac{1}{a_I}, \quad a_I^2 = 1 - \frac{r^2}{\mathbf{R}^2}, \quad \alpha_R = \frac{1}{a_R}, \quad a_R^2 = 1 - \omega^2 \sigma^2 \qquad (2.2)$$

$$\omega = \frac{a}{A^2}, \quad A^2 = r^2 + a^2, \quad \Lambda^2 = r^2 + a^2 \cos^2 \vartheta, \quad \sigma = A \sin \vartheta$$

and

$$a_T = \frac{1}{2} \left[ \left( 1 + 2\Phi_g^2 \right) \cos \eta_g - \cos \eta \right] \Phi_g^{-2}$$

$$\Phi_g^2 = \frac{r_g^2 + a^2}{r_g^2 - a^2}, \quad \cos \eta_g = \sqrt{1 - \frac{r_g^2}{\mathbf{R}^2}}, \quad \cos \eta = \sqrt{1 - \frac{r^2}{\mathbf{R}^2}} = a_I \qquad (2.3)$$

$\mathbf{R}$ and the rotational parameter a are constants. All quantities with the subscript g are the constant values of the variables at the boundary surface matching the exterior solution.

The linkage of the interior fields to the exterior fields is the junction condition

$$\mathbf{R} = A_g \sqrt{\frac{r_g}{2M}} \qquad (2.4)$$

Inserting this relation into (2.1) and (2.2) we obtain

$$a_{Ig}^2 = \frac{r_g^2 + a^2 - 2Mr_g}{r_g^2 + a^2}, \quad a_{Tg} = \cos \eta_g = a_{Ig}, \qquad (2.5)$$

which are the boundary values of the corresponding quantities of the exterior Kerr metric described in [23-27]

$$ds^2 = dx^{1^2} + dx^{2^2} + \left[\alpha_R dx^3 + i\alpha_R \omega\sigma dx^4\right]^2 + a_s^2\left[-i\alpha_R \omega\sigma dx^3 + \alpha_R dx^4\right]^2, \quad dx^1 = \alpha_s a_R dr \ , (2.6)$$

(2.1) and (2.6) have the same structure. By setting the mass parameter M=0, both metrics reduce to the same flat rotating metric, the rotation being implemented by a Lorentz transformation. So we believe that our ansatz is the natural continuation of the exterior solution into the interior region. Furthermore, we will show that both line elements are based on a similar geometrical structure.

## 3. The static metric

For a better understanding of the theory we start our investigations with a simplified form of (2.1), the static seed metric[1]. From the static line element

$$ds^2 = \alpha_I^2 a_R^2 dr^2 + \Lambda^2 d\vartheta^2 + \sigma^2 d\varphi^2 + a_I^2 dx^{4^2}, \quad dx^4 = idt \tag{3.1}$$

we read the 4-bein components and we calculate the connexion coefficients in tetrad form and we split the latter in the following manner

$$A_{mn}{}^s = B_{mn}{}^s + N_{mn}{}^s + C_{mn}{}^s + E_{mn}{}^s \tag{3.2}$$

where

$$B_{mn}{}^s = b_m B_n b^s - b_m b_n B^s, \qquad B_n = \left\{\frac{a_I}{\rho_E}, 0, 0, 0\right\}$$

$$N_{mn}{}^s = m_m N_n m^s - m_m m_n N^s, \quad N_n = \left\{0, \frac{1}{\rho_H}, 0, 0\right\}$$

$$C_{mn}{}^s = c_m C_n c^s - c_m c_n C^s, \quad C_n = \left\{a_I \frac{r}{A\Lambda}, \frac{1}{\Lambda}\cot\vartheta, 0, 0\right\} \tag{3.3}$$

$$E_{mn}{}^s = -\left[u_m E_n u^s - u_m u_n E^s\right], \quad E_n = \left\{\frac{1}{2\mathbf{R}\,\Phi_g^2 a_I a_R}\sin\eta, 0, 0, 0\right\}, \quad \sin\eta = -\frac{r}{\mathbf{R}}$$

---

1    A similar attempt for the exterior solution we have published in [26]

where $\rho_E$ is the curvature vector of the BL-ellipses and $\rho_H$ the curvature vector of their hyperbolic orthogonal trajectories

$$\rho_E = \frac{\Lambda^3}{rA}, \quad \rho_H = -\frac{\Lambda^3}{a^2 \sin\vartheta \cos\vartheta}. \tag{3.4}$$

$m_n=\{1,0,0,0\}$, $b_n=\{0,1,0,0\}$, $c_n=\{0,0,1,0\}$, $u_n=\{0,0,0,1\}$ are the orthogonal unit vectors. E is the force of gravity. From the Ricci tensor

$$R_{mn}(A) = A_{mn}{}^s{}_{|s} - A_{n|m} - A_{rm}{}^s A_{sn}{}^r + A_{mn}{}^s A_s \tag{3.5}$$

we derive

$$
\begin{aligned}
&-\left[ B_{n\,\|m} - B_{n\,\|s} b^s b_m + B_n B_m \right] - b_n b_m \left[ B^s{}_{\|s} + B^s B_s \right] \\
&-\left[ N_{n\,\|m} - N_{n\,\|s} m^s m_m + N_n N_m \right] - m_n m_m \left[ N^s{}_{\|s} + N^s N_s \right] \\
&-\left[ C_{n\,\|m} + C_n C_m \right] - c_n c_m \left[ C^s{}_{\|s} + C^s C_s \right] \\
&+\left[ E_{n\,\|m} - E_n E_m \right] + u_n u_m \left[ E^s{}_{\|s} - E^s E_s \right] = -\kappa\left( T_{mn} - \frac{1}{2} g_{mn} T \right)
\end{aligned}
\tag{3.6}
$$

We use the graded derivatives introduced in [28, 29]

$$
\begin{aligned}
m_{m\,\|n} &= m_{m|n} = 0, \quad b_{m\|n} = b_{m|n} = 0, \quad c_{m\,\|n} = c_{m|n} - B_{nm}{}^s c_s - N_{nm}{}^s c_s = 0 \\
u_{m\,\|n} &= u_{m|n} - B_{nm}{}^s u_s - N_{nm}{}^s u_s - C_{nm}{}^s u_s = 0
\end{aligned}
\tag{3.7}
$$

They transform covariantly in the lower dimensional subspaces spanned by the unit vectors and simplify the calculations considerably. Solving (3.6) by inserting (3.3) we obtain

$$\kappa \left( T_{mn} - \frac{1}{2} g_{mn} T \right) = - m_m m_n \left( M_0 B_0 + M_0 C_0 - M_0 E_0 \right)$$
$$- b_m b_n \left( M_0 B_0 + B_0 C_0 - B_0 E_0 \right)$$
$$- c_m c_n \left( M_0 C_0 + B_0 C_0 - C_0 E_0 \right)$$
$$+ u_m u_n \left( M_0 E_0 + B_0 E_0 + C_0 E_0 \right) \qquad . \quad (3.8)$$
$$- \left( m_m m_n + b_m b_n \right) \tilde{\Omega}^{s3} \tilde{\Omega}_{3s} \sin^2 \eta$$
$$+ \left( m_m m_n + c_m c_n \right) N_s C^s \sin^2 \eta + 2 N_{(m} E_{n)}$$
$$+ \left( m_m m_n + u_m u_n \right) E_s F^s$$

The new quantities

$$M_0 = - \frac{1}{\mathbf{R} a_R}, \quad B_0 = \frac{1}{\rho_E} \sin \eta, \quad C_0 = \frac{r}{A \Lambda} \sin \eta, \quad E_0 = - \frac{1}{2 \mathbf{R} \Phi_g^2 a_T a_R} \cos \eta \quad (3.9)$$

we will identify in a subsequent paper with the generalized second fundamental forms of a surface endowed with nonholonomicity.

$$\tilde{\Omega}^{s3} \tilde{\Omega}_{3s} = \tilde{H}_{13} \tilde{H}_{13} - \tilde{H}_{23} \tilde{H}_{23}, \quad \tilde{H}_{s3} = i \, \alpha_R^2 \omega \left\{ \frac{r}{\Lambda} \sin \vartheta, \frac{A}{\Lambda} \cos \vartheta, 0, 0 \right\} (3.10)$$

can be explained as contributions of the evolutes of the BL-ellipses as mentioned in [23]. We will come across the quantity

$$F_s = \alpha_R^2 \omega^2 \sigma \sigma_{|s} \tag{3.11}$$

in the next Section. From (3.8) we calculate the components of the stress-energy tensor as

$$\kappa T_{11} = B_0 C_0 - B_0 E_0 - C_0 E_0$$
$$\kappa T_{12} = E_1 N_2$$
$$\kappa T_{22} = M_0 C_0 - M_0 E_0 - C_0 E_0 - F_1 E_1 - N_2 C_2 \sin^2 \eta \qquad , \quad (3.12)$$
$$\kappa T_{33} = M_0 B_0 - M_0 E_0 - C_0 E_0 + \tilde{\Omega}^{s3} \tilde{\Omega}_{3s} \sin^2 \eta$$
$$\kappa T_{44} = M_0 B_0 + M_0 C_0 + B_0 E_0 + \tilde{\Omega}^{s3} \tilde{\Omega}_{3s} \sin^2 \eta - N_2 C_2 \sin^2 \eta$$

which are covariantly conserved.

## 4.   The Kerr interior

In the last Section we have investigated with the help of a seed metric the properties of the geometrical quantities we need to understand the proposed Kerr interior. An anholonomic intrinsic transformation

$$\Lambda_3^{3'} = \alpha_R, \quad \Lambda_4^{3'} = i\alpha_R\omega, \quad \Lambda_3^{4'} = -i\alpha_R\omega\sigma^2, \quad \Lambda_4^{4'} = \alpha_R$$
$$\Lambda_{3'}^3 = \alpha_R, \quad \Lambda_{4'}^3 = -i\alpha_R\omega, \quad \Lambda_{3'}^4 = i\alpha_R\omega\sigma^2, \quad \Lambda_{4'}^4 = \alpha_R \tag{4.1}$$

with

$$\Lambda_i^k \Lambda_j^{i'} = \delta_j^k \tag{4.2}$$

transforms the 4-bein fields and the metric as

$$\overset{m}{e}_i = \Lambda_i^{i'} \overset{m}{e}_{i'}, \quad \overset{i}{e}_m = \Lambda_{i'}^i \overset{i'}{e}_m, \quad g_{ik} = \Lambda_{ik}^{i'k'} g_{i'k'}, \tag{4.3}$$

where the primed indices denote the BL-co-ordinates of the seed metric. Applying this transformation to the tetrad connexion we obtain

$$A_{mn}^{\ \ s} = {}^*A_{mn}^{\ \ s} + G_{mn}^{\ \ s} \tag{4.4}$$

where the *A are the connexion coefficients of the seed metric and

$$G_{mns} = \left[ g_{sr} \overset{k'}{e}_m \overset{l'}{e}_n \overset{r}{e}_{j'} + g_{nr} \overset{k'}{e}_s \overset{l'}{e}_m \overset{r}{e}_{j'} + g_{mr} \overset{k'}{e}_s \overset{l'}{e}_n \overset{r}{e}_{j'} \right] \Lambda_j^{j'} \Lambda_{[l'|k']}^j \tag{4.5}$$

is the dynamical part of the Kerr interior connexion. Calculating G with the help of (4.1) we obtain

$$G_{mn}^{\ \ s} = c_m \left[ F_n c^s - c_n F^s \right] - u_m \left[ F_n u^s - u_n F^s \right] + H_{mn}^{\ \ s},$$
$$H_{mns} = -\Omega_{nm}^T u_s + \Omega_{sm}^T u_n + \Omega_{sn}^T u_m + \alpha_T D_{ns} u_m \tag{4.6}$$

wherein the new quantities are defined by

$$\Omega_{mn}^T = -\left[ H_{nm}^T + D_{nm}^T \right], \quad H_{nm}^T = a_T \left[ H_{nm} + D_{[nm]} \right], \quad D_{nm}^T = \alpha_T D_{(nm)} \tag{4.7}$$

and where

$$H_{nm} = 2 i \alpha_R^2 \omega \, \sigma_{[[n} c_{m]}$$ (4.8)

is the relativistic generalization of the Coriolis force and

$$D_{nm} = i \alpha_R^2 \omega_{|n} \sigma c_m$$ (4.9)

is the contribution from the differential rotation of the source. Shears $D_{nm}^T$ arise as neighboring layers of the fluid have different orbital velocities. The Ricci tensor has the same structure as the Ricci of the exterior solution

$$
\begin{aligned}
R_{mn} = &-\left[ N_{n\|m} - N_{n\|s} m^s m_m + N_n N_m \right] - m_n m_m \left[ N_{\|s}^s + N^s N_s \right] \\
&-\left[ B_{n\|m} - B_{n\|s} b^s b_m + B_n B_m \right] - b_n b_m \left[ B_{\|s}^s + B^s B_s \right] \\
&-\left[ C_{n\|m}^T + C_n^T C_m^T \right] - c_n c_m \left[ C_{T\|s}^s + C_T^s C_s^T - \Omega_T^{rs} \Omega_{sr}^T \right] \\
&+\left[ E_{n\|m}^T - E_n^T E_m^T \right] + u_n u_m \left[ E_{T\|s}^s - E_T^s E_s^T - \Omega_T^{rs} \Omega_{sr}^T \right] \\
&+ u_m \left[ \Omega_{Tn\|s}^s + 2 \Omega_{[ns]}^T E_T^s \right] + u_n \left[ \Omega_{Tm\|s}^s + 2 \Omega_{[ms]}^T E_T^s \right] - 2 \Omega_{m3}^T \Omega_{n3}^T
\end{aligned}
$$ , (4.10)

$$
\begin{aligned}
E_{n\|m}^T &= E_{n|m}^T - B_{mn}^{\ \ s} E_s^T - N_{mn}^{\ \ s} E_s^T - C_{mn}^{T\ \ s} E_s^T \\
C_{mn}^{T\ \ s} &= c_m C_n^T c^s - c_m c_n C_T^s \\
C_m^T &= C_m + F_m, \quad E_m^T = E_m + F_m
\end{aligned}
$$ (4.11)

The covariantly conserved components of the stress-energy tensor are

$$\kappa T_{11} = B_0 C_0 - B_0 E_0 - C_0 E_0 - \Omega^{s3}\Omega_{3s} + \Omega_T^{s3}\Omega_{3s}^T + F^s E_s$$

$$\kappa T_{22} = M_0 C_0 - M_0 E_0 - C_0 E_0 - \Omega^{s3}\Omega_{3s} + \Omega_T^{s3}\Omega_{3s}^T + N_s C^s \sin^2\eta - 2F^s E_s$$

$$\kappa T_{33} = M_0 B_0 - M_0 E_0 - C_0 E_0 - 2M_0 F_0 - 2F^s E_s +$$
$$+ \tilde{\Omega}^{s3}\tilde{\Omega}_{3s}\sin^2\eta + +\Omega^{sr}\Omega_{rs} + \Omega^{s3}\Omega_{s3} - \Omega_T^{sr}\Omega_{rs}^T - \Omega_T^{s3}\Omega_{s3}^T$$

$$\kappa T_{34} = \Omega_{03}^T M_0 + \Omega_{03}^T E_0 - \left[1 - \frac{a_T^2}{a_I^2}\right]\Omega_{13}^T B_1 - \left[1 - \frac{a_I^2}{a_T^2}\right]\Omega_{23}^T N_2 \qquad . \qquad (4.12)$$

$$\kappa T_{44} = M_0 B_0 + M_0 C_0 + B_0 C_0 + 2M_0 F_0 - N_s C^s \sin^2\eta +$$
$$+ \tilde{\Omega}^{s3}\tilde{\Omega}_{3s}\sin^2\eta - \Omega^{sr}\Omega_{rs} + \Omega^{s3}\Omega_{s3} + \Omega_T^{sr}\Omega_{rs}^T - \Omega_T^{s3}\Omega_{s3}^T$$

$$\Omega^{sr}\Omega_{rs} = 2\Omega^{s3}\Omega_{3s} = 2\left[H^{13}H_{13} - H^{23}H_{23}\right]. \qquad (4.13)$$

In addition to the above field equations the Maxwell-like equations

$$F_{[m\|n]} - 2\Omega_{T[m}^s\Omega_{n]s}^T = 0, \quad E_{[m\|n]} = 0, \quad \Omega_{[mn\|s]}^T + \Omega_{[mn}^T E_{s]}^T = 0 \qquad (4.14)$$

are satisfied. On the boundary surface the hydrostatic pressure $T_{11}$ vanishes and the matter current reduces to $\kappa T_{34} = \Omega_{03}^T M_0 + \Omega_{03}^T E_0$ where $M_0$ and $E_0$ are the curvatures of the embedded interior and exterior surfaces respectively as we will show in the subsequent paper. All the above defined field strengths coincide with the analogous field strengths of the exterior solution for $r = r_g$. Thus both the metric and the first derivatives of the metric match at the boundary surface.

The field strengths are well-behaved except at $r = 0$, $A = a$, where the BL-ellipses degenerate to a line segment fixed by the foci of the confocal BL-ellipses. On that line segment the ellipsis curvature vector $\rho_E$ is infinitely large and the corresponding field strengths are zero. Only for $\vartheta = \pi/2$, at the 'vertices' of the ellipsis, $\rho_E$ vanishes and the corresponding field strengths get infinitely large. Rotating this line segment through $\varphi$, one gets a disk with a singular rim, the well-known Kerr singularity. Setting the rotational parameter a to zero the Kerr singularity reduces to the Schwarzschild singularity at $r = 0$. The appearance of singularities of this kind is a general feature of field theories where the field strengths are of structure $1/r^n$.

## 5. Outlook

In the present paper we presented an exact solution of the Einstein field equations which could serve as Kerr interior as it matches the Kerr exterior at the boundary surface. Moreover, both solutions are on the same geometric footing. In an earlier paper we have shown that the Kerr metric is the metric of a surface embedded in a higher dimensional flat space. In a further paper we will show that the interior solution proposed in this paper can be represented by a surface, too, which matches the surface of the exterior solution.